# Optical Control of Fluorescence through Plasmonic Eigenmode Extinction

Xiaoying Xu[1], Shih-Che Lin[2], Quanshui Li[3], Zhili Zhang[3], Ilia N. Ivanov[4], Yuan Li[5], Wenbin Wang[7,1], Baohua Gu[5], Zhenyu Zhang[6], Chun-Hway Hsueh[2], Paul C. Snijders[1,7,*], Katyayani Seal[1,7,*]

[1] Materials Sciences and Technology Division, Oak Ridge National Laboratory, Oak Ridge, Tennessee 37831, United States

[2] Department of Materials Science and Engineering, National Taiwan University, Taipei 10617, Taiwan

[3] Department of Mechanical, Aerospace and Biomedical Engineering, the University of Tennessee, Knoxville TN 37996

[4] Center for Nanoscale Materials Sciences, Oak Ridge National Laboratory, Oak Ridge, Tennessee 37831, United States

[5] Environmental Sciences Division, Oak Ridge National Laboratory, Oak Ridge, Tennessee 37831, United States

[6] International Center for Quantum Design of Functional Materials (ICQD), Hefei National Laboratory for Physical Sciences at the Microscale, University of Science and Technology of China, Hefei, Anhui, 230026, China

[7] Department of Physics and Astronomy, University of Tennessee, Knoxville, Tennessee 37996, United States

*Correspondence and requests for materials should be addressed to P.C.S. (snijderspc@ornl.gov) or K.S. (kseal06@gmail.com)

**We introduce the concept of optical control of the fluorescence yield of CdSe quantum dots through plasmon-induced structural changes in random semicontinuous nanostructured gold films. We demonstrate that the wavelength- and polarization dependent coupling between quantum dots and the semicontinuous films, and thus the fluorescent emission spectrum, can be controlled and significantly increased through the optical extinction of a selective band of eigenmodes in the films. This optical method of effecting controlled changes in the metal nanostructure allows for versatile functionality in a single sample and opens a pathway to *in situ* control over the fluorescence spectrum.**

Fluorescent compounds play an important role in solid state lighting, solar energy conversion and as biomarkers in the life sciences. Controlled strategies to manipulate their fluorescent properties are therefore highly desirable. The recent successes in enhancing the magnitude of the fluorescent response of these fluorophores through proximity to metal nanostructures[1-19] represents a fundamental change in the field of optically active nanomaterials. The physics behind such metal-enhanced fluorescence (MEF) is a result of the near-field interaction between the excited fluorophore and the plasmonic response of the metal nanostructure through charge transfer[20] or electromagnetic interactions.[21] It provides a unique opportunity for control of the fluorophore emission properties. However, the necessary geometric optimization (size, shape, and arrangement) of the vicinal metal nanostructures in this approach also introduces an additional level of complexity.[5,10,11,14] Even though increasing attention is now being given to such plasmon-induced spectral reshaping,[21-26] there are no studies to date that attempt control over the spectral profile of fluorophores through *photo-induced* plasmon-mediated changes to the metal nanostructures, despite the distinct advantages that are inherent to this approach.

In this work, we present a proof of principle experiment demonstrating the concept of optical control of the fluorescence spectrum of CdSe quantum dots (QDs) through plasmon-induced changes in proximal nanostructured gold semicontinuous films, see Fig. 1(a). Semicontinuous metal films are alluring for their broad distribution of finely spaced eigenmodes with wide-ranging applications from enhanced Raman sensors to near-perfect absorbers.[27] We utilize the unique advantage of their broad absorption spectrum which renders them amenable to optical modification by selective eigenmode extinction.[28,29] The optically customized distribution of surface plasmon modes in turn modifies the emission spectrum of the fluorophores that are vicinal to the films. Our results demonstrate such optical control of fluorescent emission spectra without involved nanostructure design and modeling. They also establish the role of the spectral density of plasmonic eigenmodes and their interaction with fluorophores. Moreover, the approach of plasmon-mediated control provides potential for unprecedented functional versatility from a single sample. This versatility is enabled by the capacity of the semicontinuous films to support multiple spectral bands within the same spatial location: apart from controlling the wavelength where the optical properties of the films can be tailored, each spectral feature can also be designed to be polarization sensitive. This allows for polarization driven switching of fluorescence and opens a pathway to *in situ* dynamic control over the fluorescence properties.

**Methods**

Semicontinuous gold films of nominal thickness 3 nm, 5 nm and 6 nm were grown on (001) quartz substrates by thermal evaporation (Figure 1(a)). Multiple samples were characterized to ensure repeatability. The absorption spectra for the films were measured with a CARY5000 UV-vis spectrophotometer. Subsequently, the films were "photomodified": a linearly polarized laser source (Continuum Surelite I-10 Nd:YAG laser pumped ND6000 dye laser operating at 670 nm, fluence 10 mW/cm$^2$) was used to burn a "spectral hole" in the otherwise broad absorption spectrum.[28,29] Figure 1(b) is a schematic of the spectroscopic approach, illustrating absorption and fluorescence spectra before and after photomodification, as well as their overlap. The spectral hole in the absorption spectrum after photomodification is caused by the nanoscale sintering of Au-metal clusters supporting specific highly resonant plasmonic eigenmodes. This sintering results in the extinction of these modes, and consequently a minimum (the spectral hole) is formed in the absorption spectrum at the chosen wavelength and polarization of the irradiating source. Subsequently, a spacer layer of 25 nm thick $SiO_2$ was deposited to reduce fluorescence quenching while retaining effective plasmon fluorophore coupling[18] as well as to protect the semicontinuous film surface (Fig. 1(a)). Approximately 10 uL of an aqueous solution (4mg/ml) of monodisperse CdSe quantum dots of radius 5 nm (AmericanElements) was then deposited on the $SiO_2$ surface and dried (spot size ~3 mm) prior to the measurements. The QD emission spectrum was designed to overlap with the region of high absorption (as well as the post-modification absorption minimum) in the unmodified Au film spectrum (see Figure 1(b)), to ensure maximum coupling with plasmon modes in the film. The broad emission curve of the QDs, attributed to inter-dot coupling,[30,31] is advantageous since it allows for a larger number of available modes in the QD system. These modes couple with the eigenmodes of the semicontinuous films, leading to a larger potential for modification of the fluorescence emission. For the spectral response a diode laser at 532 nm (intensity=5mW) was made incident on the samples and the fluorescent emission was recorded for forward transmitted geometry at an angle of 135 degrees to the surface normal (Figure 1(a)) with a Princeton Instruments SpectraPro 2300i monochromator with a ICCD 7397-0035 detector. The sample was rotated in-plane between two perpendicular orientations so that the excitation polarization was either parallel or perpendicular to the polarization orientation used during photomodification. Figure 1(b) shows the relative spectral positions of the pre- and post-photomodification peaks of the Au film, and the emission peaks of the CdSe quantum dots. AFM pictures of the semicontinuous Au film in Fig. 1(c) show an increase in grain size with increasing thickness. The RMS roughness values were 0.92, 2.3 and 2.0 nm respectively for the 3 nm, 5 nm and 6 nm samples.

**Results and discussion**

Figure 2(a) shows the absorption spectra using *s*- and *p*-polarized light for the 5 nm thick film. It is well-known[28,32,33] that the photomodification process is strongly polarization sensitive: a spectral hole only appears in the measured spectrum of a modified film when it is measured using the same polarization as was used to modify the film. Indeed, the absorption spectrum (absorption≈1-transmission) of a film that was modified using *p*-polarized light reveals a reduced absorption near 750 nm when measured using *p*-polarized light, whereas no spectral hole is observed on the same sample when measuring using *s*-polarized light. This allows the use of the spectrum measured using *s*-polarized light as the unmodified reference spectrum, as has been reported in Refs. [28,32,33]. It also eliminates possible sample variations by allowing the modified and unmodified spectra to be obtained from the very same sample area.

Apart from the reduction in absorption near 750 nm, other parts of the spectrum are essentially unchanged. This indicates that the photomodification occurred just above the photomodification threshold. In such near-threshold photomodification the film morphology undergoes subtle changes induced by localized sintering of a few metal nanoparticles.[32,33] Photomodification at fluences significantly above the photomodification threshold (enabled via an increased number of pulses or increased laser intensity) results in significantly larger reductions in the absorption spectrum. Significant changes in the macroscopic morphology of the film will then occur due to the melting and coalescence of multiple metal islands.[29] This would impose substantial changes to the surface plasmon modes supported by the film, and the correlation between the unmodified and the modified metal films would be lost as an essentially new metal film with a different morphology would be created. Instead, in order to study the influence of the coupling between the plasmonic eigenmodes and the fluorophores, we aimed for optimal photomodification conditions just above the photomodification threshold.[32,33] The signature of this limited modification is the burning of a spectral hole in our absorption spectrum without structural changes on mesoscopic and macroscopic scales. Indeed, AFM images of the Au semicontinuous film before and after photomodification reveal no large scale morphological changes, as illustrated for the 5 nm film in Fig. 2(b). The average grain structure and size appear unaltered upon photomodification, while the RMS roughness before and after photomodification was 2.2 and 2.1 nm, respectively. This is within the precision of our AFM measurements, corroborating that the photomodification occurred slightly above the photomodification threshold.[28,29]

The fluorescence spectra show a similar dependence on the polarization as the absorption spectra, see the inset in Fig 2(a) for the fluorescence spectra of the CdSe dots on a 5 nm thickness film. The unmodified fluorescence spectrum (black) is perfectly replicated by the *s*-polarized spectrum (blue) whereas the *p*-orientation (red) shows a modified spectrum with a higher intensity of transmitted fluorescent emission. This behavior, somewhat unexpected because fluorescence is a secondary process and the photomodification was carried out before the QDs were deposited, is similar to the polarization selectivity in the absorption spectra discussed above. This similarity conveniently allows us to use the data measured using *s*-polarized light as the (un-photomodified) reference for both absorption[28,29] as well as fluorescence experiments. Notably, this change in polarization-dependent fluorescence amplitude demonstrates an *in situ* effective deactivation of the photomodification in the fluorescence spectrum by (dynamic) switching from *p*- to *s*- polarized light. A comparison between the normalized spectra with *p* and *s* orientations (red and black, respectively, in the inset of Figure 2(c)) for the 5 nm sample shows a red shift in the fluorescence for the *p* orientation. This net fluorescence emission peak shift upon photomodification is plotted as a function of thickness in Figure 2(c). The data show a maximum red shift (12 nm) for the 5 nm sample, with the 6 nm sample showing a blue shift after photomodification. The normalized fluorescence emission spectra of unmodified films with a different thickness are identical (black and blue curves, inset Fig. 2(c)). Therefore, the spectral *reshaping* of the fluorescence emission is not due to quenching or thickness related differences in the statistical distribution of Au particles. We attribute the spectral reshaping predominantly to changes in the complex interaction between the QDs and the eigenmodes of the semicontinuous film due to the photomodification process.

To quantify the change in the optical response of the films due to photomodification, the ratio of the unmodified to modified absorption spectrum of the bare metal film (the photomodification ratio, $A(\lambda)/A_{mod}(\lambda)$) was obtained for the three thicknesses shown in Figure 3(a). This ratio peaks at a wavelength slightly greater than the wavelength of photomodification (~670 nm). For the 5 nm and 6 nm samples $A(\lambda)/A_{mod}(\lambda)$ peaks at a wavelength of ~740 nm whereas for the 3 nm sample it peaks at 710 nm.

The absorption photomodification ratio is highest for the 5 nm sample. For thicknesses away from this value it decreases in magnitude.

The local sintering in the photomodification process results in a reduction in the number of eigenmodes that absorb at the photomodification wavelength (and polarization),[32,33] and couple to the fluorophores. Hence one expects the fluorescence modification to be correlated with the absorption modification. The fluorescence modification ratio shown in Figure 3(b) was obtained by dividing the fluorescence emission from the modified metal-QD composite system by the unmodified emission for each of the three thicknesses ($F_{mod}(\lambda)/F(\lambda)$). The significantly asymmetric shape of the fluorescence modification ratios demonstrates that the extinction of a band of eigenmodes not only changes the fluorescent intensity, but in fact reshapes the fluorescent emission spectrum. The fluorescence ratio shows a thickness-dependence and is spectrally broadened with increasing metal content because of the spectral broadening in the absorption spectrum (Fig. 1(c)). This indicates that a larger fraction of the extended plasmonic eigenmodes (eigenmodes with larger localization lengths that exist at increased metal content[34,35]) can interact with the QDs resulting in a broadened coupled emission spectrum with increased metal content. The number of eigenmodes that is affected by the photomodification also increases with increasing metal concentration up to the percolation threshold, where not only the population of eigenmodes with a wide distribution of localization lengths is largest, but the scattering strength is also the highest.[34,35] The broad fluorescence modification ratio of the 5 nm sample therefore suggests that the it is closest to the percolation threshold. As a result, decreasing the absorption (due to the spectral hole) in the 5 nm sample will lead to the largest change in fluorescence intensity in a broader wavelength region. At lower metal concentrations, a smaller number of modes with smaller localization lengths and higher local field intensities dominate the near-field intensity statistics.[34,35] The overall result is a reduced coupling to the fluorophores as evident from the narrower and smaller change in the fluorescence emission spectrum observed for the 3 nm film as compared to the 5 nm film (Fig 3(b)). Note that the fluorescence emission for the 5 nm film increases with a factor 1.6 upon photomodification (Fig. 3(b)), while the absorption decreases only by approximately 22% (Figs. 2(a) and 3(a)). Moreover, for the 3 nm and 6 nm films, the change in fluorescence is significantly smaller than for the 5 nm film despite comparable changes in the absorption. This difference between the thickness dependence of the fluorescence and absorption spectra demonstrates that the decreased absorption is not the dominant cause for the increase in fluorescence upon photomodification. This is further confirmed by the fact that the fluorophore emission spectra for unmodified films with different thicknesses (and therefore varying absorption) are identical, see the overlapping black and blue curves in the inset in Fig. 2(c). Therefore, the observed changes in the fluorescence spectra are not solely due to changes in absorption of the film. Instead, they are predominantly due to changes in a complex interaction between QDs and the eigenmodes of the semicontinuous film upon photomodification.

Our results indicate that the extinction of a band of eigenfrequencies by photomodification has a profound impact on metal-fluorophore coupling. Although the spectral shift is modest, it is reproducible and serves conclusively as a proof of principle utilizing optical modification of specific eigenmodes as a means of tailoring the fluorescence response. The thickness trend of the spectral shift confirms that the extinction of the extended or delocalized modes associated with metal concentrations closer to percolation (i.e. the 5 nm sample) has a stronger effect on the active fluorophore modes detected in the measurement than the extinction of modes with smaller localization lengths at metal concentrations above percolation.[35] The impact of the coupling between metal and fluorophore is also demonstrated by the experimental polarization dependence of the fluorescence (inset in Fig. 2(a)) indicating that the fluorescent spectrum is principally governed by the optical response of the metal nanostructure. The emitted fluorescence takes on the polarization dependence properties of the eigenmodes of the semicontinuous film, thus demonstrating effective optical manipulation of the fluorescent emission spectrum.

## Conclusions

These results are a proof of principle demonstration that optical tailoring of the nanostructure in metallic composites is a viable way of controlling the fluorescence emission spectrum of fluorescent species. Specifically, random metal-dielectric films provide an appealing avenue for plasmon-induced tailoring of fluorescent emission by modifying the metal optical response through the process of photomodification. The optical extinction of a selective band of eigenmodes in the semicontinuous film can reshape and increase the fluorescence emission spectrum. The thickness dependence of our results suggests that to maximize the magnitude of the change, coupling with a broad distribution of plasmonic eigenmodes with enhanced scattering strength is more advantageous. Instead, at lower metal concentrations with a lower absorption and a narrower distribution of eigenmodes, there is the potential for more subtle (spectrally narrower) changes in the fluorescence emission upon photomodification.[34,35] The fact that multiple spectral holes may be burned in a single area through photomodification[28,29] provides for unprecedented functional versatility through the potential to use multiple fluorophores and excitation lasers on the same sample area, thus circumventing the involved post-synthesis nanostructure modification used in prior studies.[21-26] Moreover, as demonstrated here, the highly polarization-sensitive nature of each spectral hole allows for *in situ* polarization driven dynamic switching for nanoscale light sources that can be spectrally tailored. We envision that the optical control of the fluorescent emission spectrum demonstrated here can be enhanced by careful tuning of laser excitation wavelengths, polarizations and, photomodification of semicontinuous films. The results of plasmon controlled fluorescent emission presented here have the potential to significantly impact a broad range of potential applications for fluorescent molecules as narrow width nanoscale light sources, ultra-bright probes with plasmon controlled fluorescence, single molecule detection, solar energy harvesting and as components for optical circuitry.

## Acknowledgements

This effort was supported by the U.S. Department of Energy (DOE), Office of Science, Basic Energy Sciences (BES), Materials Sciences and Engineering Division (XX, PCS, KS). WW and ZZ were supported respectively by DOE BES DE-SC0002136 and DE_ER45958. BG and YL were supported by the Laboratory Directed Research and Development Program of Oak Ridge National Laboratory (ORNL), managed by UT-Battelle, LLC, for the U.S. DOE, and characterized the quantum dots and performed initial fluorescence measurements. A portion of this research was conducted at the Center for Nanophase Materials Sciences, which is sponsored at ORNL by the Scientific User Facilities Division, Office of BES, U.S. DOE (INI). We acknowledge partial funding support from Ministry of Science and Technology, Taiwan under Contract number MOST 103-2221-E-002-076-MY3 (SCL, CH).

## Author contributions

Sample fabrication was done by XX, QL, ZhiliZ, YL and KS. AFM characterization was performed by WW. Absorption and fluorescence measurements were performed by XX and KS with help from INI, YL, and BG. Data analysis and interpretation was done by XX, PCS and KS, aided by discussions with SCL, ZhenyuZ and CH. XX, PCS and KS wrote the paper. All authors commented on the manuscript.

## Competing financial interests

The authors declare no competing financial interests.

Figure legends

Fig. 1. (a) Experimental geometry. (b) Schematic of the spectroscopic approach: pre- and post-photomodification absorption and fluorescence spectra. The spectroscopic experiment was set up by choosing the spectral region of photomodification to overlap with the fluorescent emission spectrum of the quantum dots. (c) Absorption at three values of Au mass thickness. Insets in (c) are the corresponding 800 nm ×800 nm AFM images.

Fig. 2. Au mass thickness 5nm. (a) Pre- (*s*-polarization) and post-photomodification (*p*-polarization) absorption spectra of the 5 nm sample. Inset: fluorescence emission spectra demonstrating polarization selectivity upon photomodification as the unmodified (black) spectrum is identical to the spectrum measured using *s*-polarized light, while measuring using *p*-polarized light reveals photomodification changes. (b) AFM images of the 5 nm film before (top) and after (bottom) photomodification, demonstrating the absence of large scale structural changes. Black scale bars are 1 μm. (c) Fluorescence peak shift as a function of metal thickness. Inset: normalized fluorescence emission. Spectral reshaping upon photomodification is illustrated with the red and black spectra for the 5 nm sample. The unmodified fluorescent emission spectra for different thicknesses of 5 nm (black) and 3 nm (blue) samples in the inset are identical.

Fig. 3. Absorption modification ratios $A(\lambda)/A_{mod}(\lambda)$ (a) and fluorescence modification ratios $F_{mod}(\lambda)/F(\lambda)$ (b) for the three Au film thicknesses.

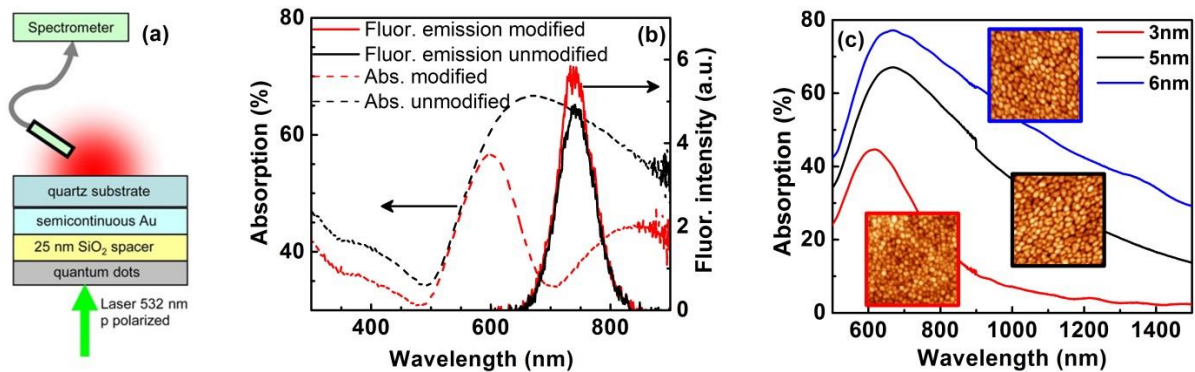

Fig. 1

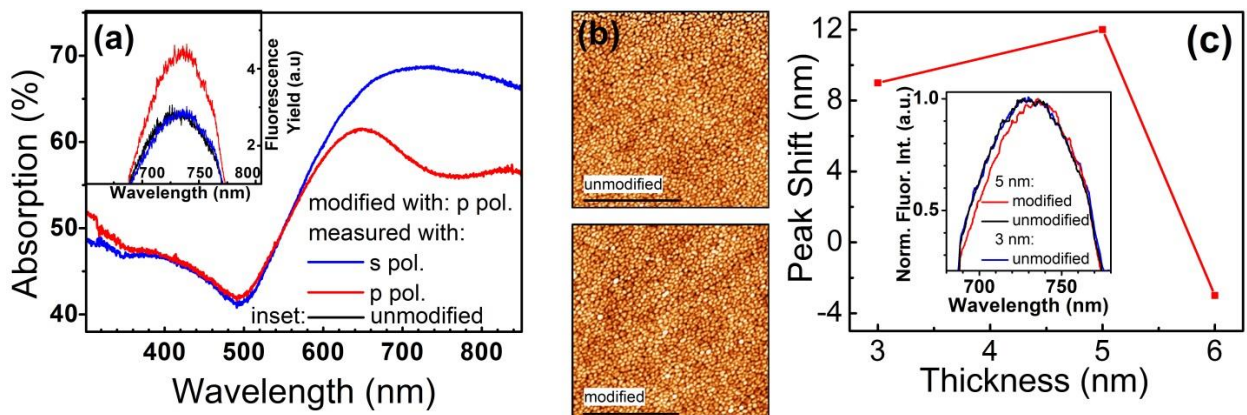

Fig. 2

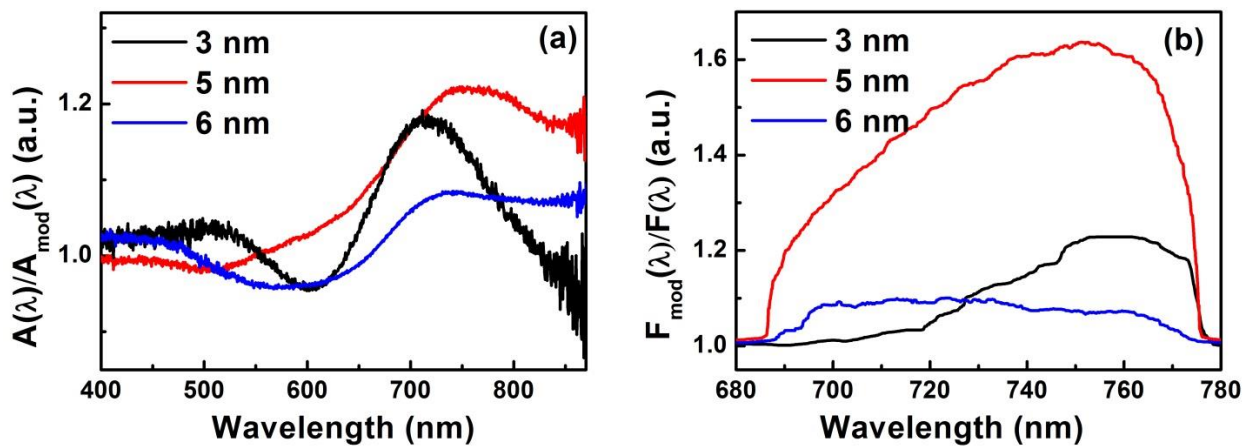

Fig. 3